\documentclass[preprint,12pt]{elsarticle}
\usepackage{amssymb}
\usepackage{amsthm}
\usepackage{lineno}
\usepackage{ragged2e}
\usepackage{hyperref}
\makeatletter
\def\ps@pprintTitle{%
  \let\@oddhead\@empty
  \let\@evenhead\@empty
  \def\@oddfoot{\footnotesize Manuscript accepted by the \textit{Chinese Journal of Physics} \hfill February 6, 2025}%
  \let\@evenfoot\@oddfoot}
\makeatother
\begin{document}

\begin{frontmatter}

\title{A Thermodynamic Model for Dark Energy Including Particle Creation or Destruction Processes}

\author[1]{José Medeiros da Costa Netto}
    \ead{josemedeiros.netto@ufpe.br}
\author[1]{Heydson Henrique Brito da Silva}
    \ead{heydson.henrique@ufpe.br}
    \affiliation[1]{
            organization={Núcleo de Formação Docente, Centro Acadêmico do Agreste, Universidade Federal de Pernambuco}, 
            city={Caruaru},
            state={Pernambuco},
            postcode={55014-900}, 
            country={Brasil}
            }
\begin{abstract}
    Thermodynamic analyses of dark energy as a relativistic fluid indicates that this intriguing component of the universe mimics a bulk viscous pressure when the parameter of its barotropic equation of state varies with time. Since in cosmology bulk viscosity and creation or destruction of matter are closely linked processes, we propose in this work a brief thermodynamic study of dark energy considering that particles can be created or destroyed in the fluid. We derive new expressions for quantities such as particle density, entropy density etc. that have been shown to be sensitive to this new ingredient. We also obtain new thermodynamic constraints and compare them with those where the number of particles is conserved. In particular, we found that in the presence of a sink, dark energy tends towards the cosmological constant over time regardless of the sign of its chemical potential and without violating the laws of thermodynamics.
\end{abstract}

\end{frontmatter}
\section{Introduction} \label{intro}

Since 1998 we have known, thanks to the independent efforts of High-z Supernova Search Team (HSST) and Supernova Cosmology Project (SCP), that there is a extra source of energy filling almost $70\%$ of the universe and that is responsible for cosmic acceleration \cite{riess, perlmutter, aghanim}. This discovery is, arguably, the most paradigmatic in cosmology in the last $25$ years because tells not only that the expansion of the universe is accelerated, but also that the agent that causes it does not interact with light. According to General Relativity, this invisible component is the dark energy. It has negative pressure by the Friedmann equations \cite{weinberg1972, weinberg2008} and it has the cosmological constant $\Lambda$ as a particular case - as it should be (see \cite{carroll, peebles}). We can model dark energy using field theory, by associating it with a quintessence scalar field, or by considering it an exotic relativistic fluid and applying the entire basis of thermodynamics to it. We use the second perspective in here.

Early works considered dark energy to be a perfect relativistic fluid with constant barotropic equation of state (EoS) and null chemical potential \cite{lima2004}. With these assumptions, they found that this component heats during the adiabatic cosmic expansion and that the phantom regime (dark energy breaking the $\Lambda$ barrier) is forbidden by the positivity of entropy. The only possible way to allowed it in this scenario is to consider negative absolute temperatures \cite{gonzalez2003, sami, gonzalez2004.1, gonzalez2004.2, babichev, myung, saridakis} (see also \cite{nojiri2003.1, nojiri2003.2, cline, samart}). However, if the chemical potential is nonzero, it is possible to achieve the phantom behavior without appeal to this unusual assumption. In fact, in \cite{lima2008, pereira} it is shown that the phantom behavior is possible when dark energy have a negative chemical potential. Not only that, but all the results saw in \cite{lima2004} are retrieved when we zero this quantity. 

More recently, some researches have studied the thermodynamic behavior of dark energy with variable EoS parameter, 

\begin{equation} \label{eq1}
    \omega(a)=\frac{p_{x}}{\rho_{x}}=\omega_{0}+\omega_{a}f(a),
\end{equation}

\noindent where $a=a(t)$ is the scale factor, $p_{x}$ and $\rho_{x}$ are the pressure and energy density of dark energy, respectively, $\omega_{0}$ and $\omega_{a}$ are constants and $f(a)$ is a function that includes all the relevant parameterizations present in the specialized literature \cite{barboza}. Assuming initially a zero chemical potential, the authors of \cite{silva2012} have shown that the fluid mimics a bulk viscous pressure $\Pi=\omega_{a}f(a)\rho_{x}$ when Eq.(\ref{eq1}) is used. Therefore, this strange component leaves the adiabatic limit and begins to generate entropy due to the bulk viscosity process. Very restrictive thermodynamics bounds were, then, imposed to $\omega(a)$. Later, these same authors included the chemical potential in the analysis \cite{silva2013}, achieving the most general model for dark energy thermodynamics known so far, from which the previous models appeared as particular cases of it.

The one thing common to all papers cited above is that they consider that the particle number of the fluid remains always constant during cosmic evolution. Matter creation or destruction is a well known process in cosmology \cite{gunzig, prigogine1988, prigogine1989}. There is an intimate relationship between this process and bulk viscosity. In fact, several studies shows that bulk viscosity implies the creation or destruction of matter \cite{calvao, lima1992, zimdahl1993, lima1996, zimdahl1996}. Then, considering that dark energy experiences a bulk viscous pressure due to the time dependent EoS parameter, it is reasonable to assume that there is a source or sink of particles in the fluid.

In this work we propose a simple type of generalization of \cite{silva2013} considering the presence of particle creation or destruction processes. First we re-derive some of the fundamental thermodynamic equations that describes dark energy taken 

\begin{equation} \label{eq2}
    N^{\mu}_{\hspace{0.5ex};\mu}=\psi=n\Gamma,
\end{equation}

\noindent where $N^{\mu}$ is the fluid particle current, $\psi$ is the source ($\psi>0$) or sink ($\psi<0$),  $n$ is the particle density and $\Gamma$ is the rate of particle creation ($\Gamma>0$) or destruction ($\Gamma<0$). This $\Gamma$ factor may or may not be constant, depending, lets us say, in first order with $a$. Nevertheless, we do not consider a specific form for $\Gamma$ and because of that we claimed this work to be a first approach. Subsequently, we determined new thermodynamic constraints that are more restrictive than those imposed in models with $\Gamma=0$, analyzing them in theoretical graphs and schematizations. We will see that the results obtained here are compatible with those works cited above when the dark component preserves the number of particles.

Throughout the whole paper we use the subscript 0 to denote quantities measured in present time, $t=t_{0}$, and consider units where $8\pi G=c=1$. Our metric signature is (+, -, -, -).
\section{Dark Energy Thermodynamics with Particle Creation or Destruction} \label{des1}

As is standard, we consider a flat, homogeneous and isotropic universe (a universe that obeys the cosmological principle) that satisfies the Friedmann-Robertson-Walker metric \cite{weinberg1972, weinberg2008}:

\begin{equation} \label{eq3}
    ds^{2}=dt^{2}-a^{2}(t)\left[dr^{2}+r^{2}d\theta^{2}+r^{2}\sin^{2}(\theta)d\phi^{2}\right].
\end{equation}

\noindent We consider dark energy as a relativistic fluid with stress-energy tensor $T^{\mu\nu}$, particle current $N^{\mu}$ and entropy current $S^{\mu}$ given respectively by

\begin{equation} \label{eq4}
    T^{\mu\nu}=\rho_{x}U^{\mu}U^{\nu}-p_{x}h^{\mu\nu},
\end{equation}

\begin{equation} \label{eq5}
    N^{\mu}=nU^{\mu},
\end{equation}

\begin{equation} \label{eq6}
    S^{\mu}=sU^{\mu}=n\sigma U^{\mu},
\end{equation}

\noindent where $\vec{U}$ is the 4-velocity, $h^{\mu\nu}\equiv g^{\mu\nu}-U^{\mu}U^{\nu}$ is the projection tensor and $s=n\sigma$ is the entropy density, with $\sigma$ being the specific entropy. The conservation of energy states that:

\begin{equation} \label{eq7}
    U_{\mu}T^{\mu\nu}_{\hspace{1ex};\nu}=\dot{\rho_{x}}+3\left(\rho_{x}+p_{x}\right)\frac{\dot{a}}{a}=0.
\end{equation}

\noindent Using Eq.(\ref{eq1}) and applying it in Eq.(\ref{eq7}), we have

\begin{equation} \label{eq8}
    \dot{\rho_{x}}+3\left(\rho_{x}+p_{0}\right)\frac{\dot{a}}{a}=-3\Pi\frac{\dot{a}}{a},
\end{equation}

\noindent where $p_{0}=\omega_{0}\rho_{x}$ and $\Pi=\omega_{a}f(a)\rho_{x}$. This result shows us that dark energy mimics a bulk viscous pressure $\Pi$ due to its time dependent EoS parameter $\omega(a)$ \cite{silva2012,silva2013}. Then, $p_{x}=p_{0}+\Pi$ is it the total pressure and $p_{0}$ is the equilibrium pressure. Therefore, dark energy is not a perfect relativistic fluid, but a imperfect one suffering from bulk viscosity. This process generates entropy in such a way that, if 

\begin{equation} \label{eq9}
    N^{\mu}_{\hspace{0.5ex};\mu}=0,
\end{equation}

\noindent then

\begin{equation} \label{eq10}
    S^{\mu}_{\hspace{0.5ex};\mu}=n\dot{\sigma}.
\end{equation}

In cosmology, creation or destruction of matter and bulk viscosity are closely linked processes \cite{calvao, lima1992, zimdahl1993, lima1996, zimdahl1996}. Since dark energy has a bulk viscous pressure as seen, it is reasonable to consider that there is a particle source or sink present in it. In this sense, Eq.(\ref{eq9}) becomes Eq.(\ref{eq2}),

\begin{equation} \label{eq11}
    N^{\mu}_{\hspace{0.5ex};\mu}=\psi=n\Gamma.
\end{equation}

\noindent When $\Gamma>0$, there is a source; when $\Gamma<0$, there is a sink, as stated in Section \ref{intro}. The divergence of the entropy current changes to:

\begin{equation} \label{eq12}
    S^{\mu}_{\hspace{0.5ex};\mu}=n\dot{\sigma}+n\sigma\Gamma.
\end{equation}

\noindent The inclusion of $\Gamma$ in nothing affects the stress-energy tensor of dark energy, so the continuity equation, Eq.(\ref{eq8}), remains the same. 

Before we continue, it is necessary to discuss whether the processes of creation or destruction of particles are adiabatic or not. In the works cited in the previous paragraph, all the proposed models aim to describe the cosmological fluid as a whole and consider, for reasons of symmetry, $\dot{\sigma}=0$. This implies that only creation processes can occur without breaking the second law of thermodynamics \cite{gunzig, prigogine1988, prigogine1989}. However, in \cite{silva2013} we see that this consideration leads to the conservation of the entropy current, which is a direct contradiction to the conclusion that there is a bulk viscous pressure mimetized by the dark energy fluid. In our research we take $\dot{\sigma}\neq0$ and because of that we do not restricted ourselves to particle creation only: destruction is also possible, and this does not break any thermodynamics law. Actually, as we will see in Section \ref{des2}, it is more probable that particle destruction take place in this exotic component of the universe.

We verified that the temperature evolution law is exactly the same as if $\Gamma=0$ \cite{silva2013}:

\begin{equation} \label{eq13}
    \frac{\dot{T}}{T}=-3\left(\frac{\partial p_{x}}{\partial \rho_{x}}\right)_{n}\frac{\dot{a}}{a},
\end{equation}

\begin{equation} \label{eq14}
    \frac{\dot{T}}{T}=-3\left(\frac{\partial p_{0}}{\partial \rho_{x}}\right)_{n}\frac{\dot{a}}{a}-3\left(\frac{\partial \Pi}{\partial \rho_{x}}\right)_{n}\frac{\dot{a}}{a}.
\end{equation}

\noindent From this and Eqs.(\ref{eq8}, \ref{eq11}), we get, respectively, the following relations for the dark energy density $\rho_{x}$, particle density $n$ and temperature $T$:

\begin{equation} \label{eq15}
    \rho_{x}=\rho_{0}\left(\frac{a}{a_{0}}\right)^{-3}\exp\left[-3\int\frac{\omega(a)}{a}da\right],
\end{equation}

\begin{equation} \label{eq16}
    n=n_{0}\left(\frac{a}{a_{0}}\right)^{-3}\exp\left(\int\Gamma dt\right),
\end{equation}

\begin{equation} \label{eq17}
    T=T_{0}\left[-3\int\frac{\omega(a)}{a}da\right].
\end{equation}

\noindent To obtain a similar expression for the entropy density $s$, lets us begin with the Euler relation for thermodynamics \cite{reichl}, 

\begin{equation} \label{eq18}
    T(n\sigma)=\rho+p-\mu n,   
\end{equation}

\noindent where $\mu$ is the chemical potential. Assuming that the quantity $\mu/T$ remains constant \cite{lima2008, pereira, silva2013}, the expression for $\sigma$ becomes:

\begin{equation} \label{eq19}
    \sigma=\frac{\rho_{0}}{T_{0}n_{0}}\left[1+\omega(a)\right]\exp\left(-\int \Gamma dt\right)-\frac{\mu_{0}}{T_{0}}.
\end{equation}

\noindent This lead to, by Eqs.(\ref{eq6}, \ref{eq12}):

\begin{equation} \label{eq20}
    s=\left\{s_{0}+\frac{\rho_{0}}{T_{0}}\left[\omega(a)-\omega_{0}\right]-\frac{\mu_{0}n_{0}}{T_{0}}\left[\int\Gamma\exp\left(\int\Gamma dt\right)dt\right]\right\}\left(\frac{a}{a_{0}}\right)^{-3}.
\end{equation}

\noindent Note that all our results recovers the ones presented at \cite{silva2013} when $\Gamma=0$, as one should expect.
\section{New Thermodynamics Constraints} \label{des2}

The total entropy of the dark energy component is given by $S=\left(n\sigma\right)V$, where $V=V_{0}\left(a/a_{0}\right)^{3}$ is the comoving volume \cite{weinberg1972, weinberg2008}. In our model, then, we have:

\begin{equation} \label{eq21}
    S=\left\{\rho_{0}\left[1+\omega(a)\right]-\mu_{0}n_{0}\exp\left(\int \Gamma dt\right)\right\}\frac{V_{0}}{T_{0}}.
\end{equation}

\noindent If there is no creation or destruction of particles happening, the expression for $S$ tends to the standard one \cite{silva2013}. From the positivity of entropy, $S\geq0$:

\begin{equation} \label{eq22}
    \omega(a)\geq\omega(a)_{min}=-1+\frac{\mu_{0}n_{0}}{\rho_{0}}\exp\left(\int \Gamma dt\right).
\end{equation}

\begin{figure}[t]
    \centering
    \includegraphics[width=0.69\textwidth]{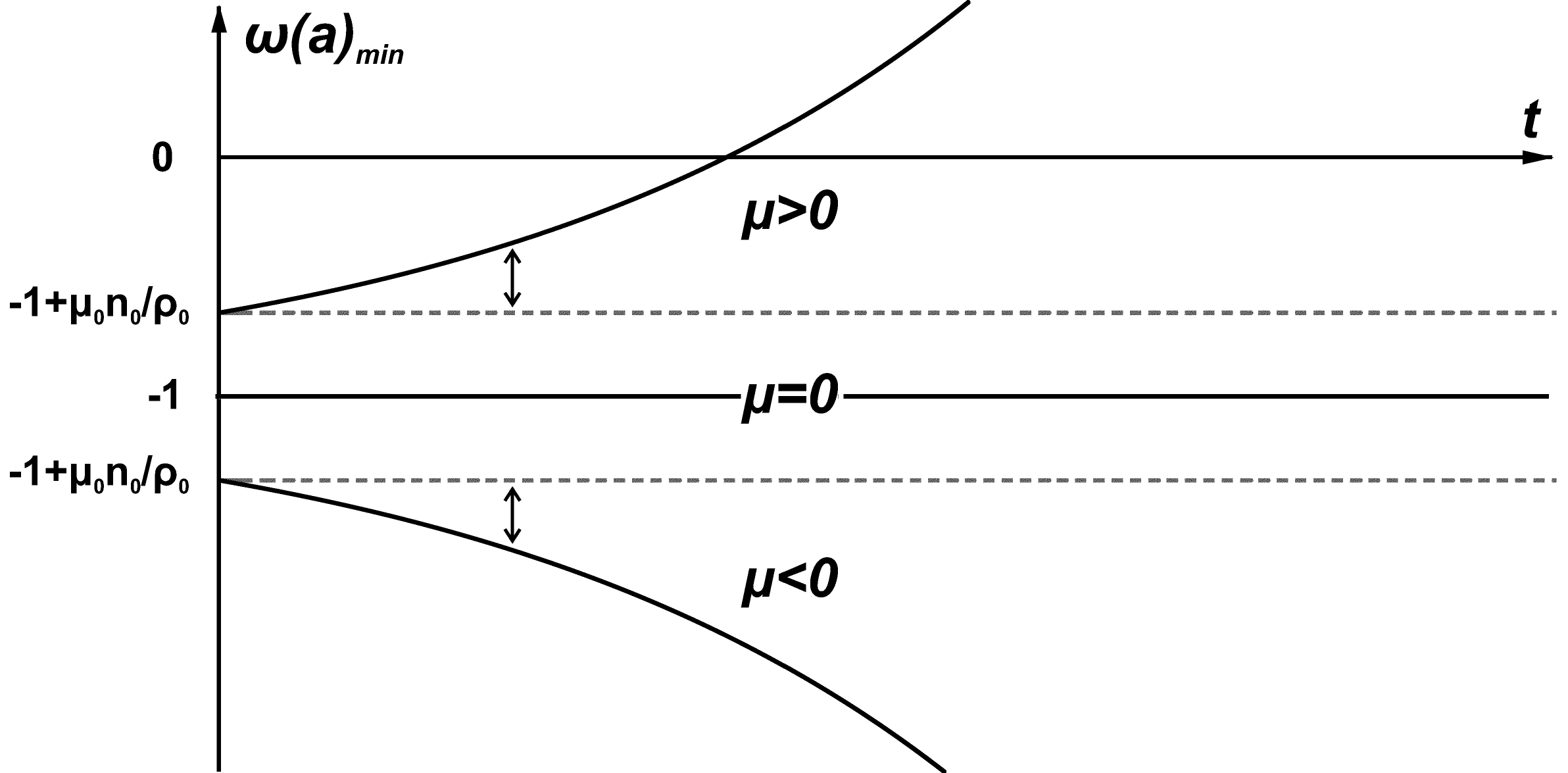}
    \caption{\justifying Theoretical representation of the behavior of Eq.(\ref{eq22}) bound for $\mu>0$, $\mu=0$ and $\mu<0$ when $\Gamma>0$. The dotted lines represent the constant values of $\omega(a)_{min}$ when there is no particle creation in the fluid.}
    \label{fig1}
\end{figure}

\begin{figure}
    \centering
    \includegraphics[width=0.69\textwidth]{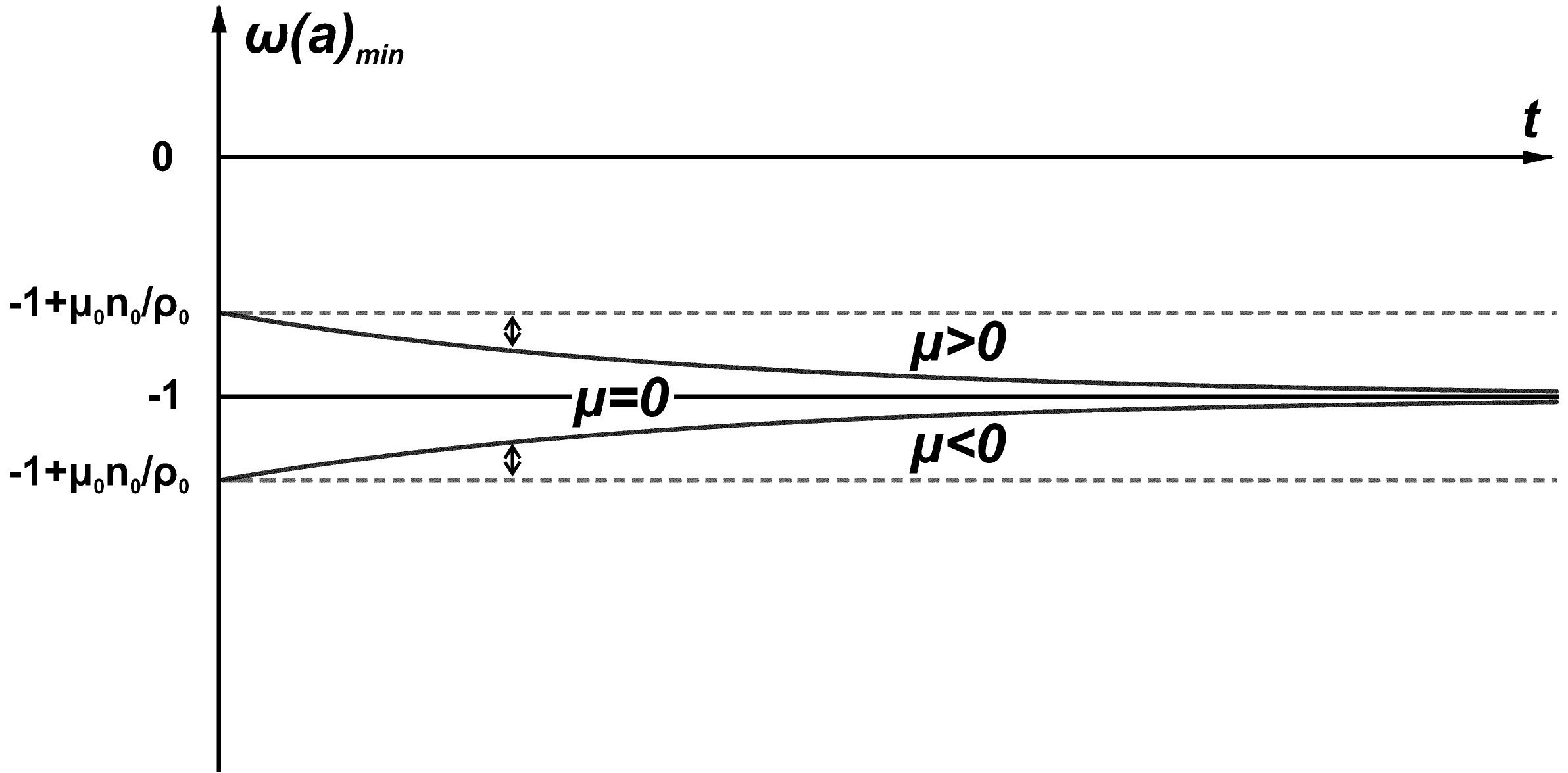}
    \caption{\justifying Theoretical representation of the behavior of Eq.(\ref{eq22}) bound for $\mu>0$, $\mu=0$ and $\mu<0$ when $\Gamma<0$. The dotted lines represent the constant values of $\omega(a)_{min}$ when there is no particle destruction in the fluid.}
    \label{fig2}
\end{figure}

\noindent As we can see, the chemical potential still plays the fundamental role of determine the interval of possible values for the EoS parameter: if $\mu>0$, then $\omega(a)_{min}>-1$ and the phantom regime is not allowed; if $\mu=0$, then $\omega(a)_{min}=-1$ and we have cosmological constant; if $\mu<0$, then $\omega(a)_{min}<-1$ and the phantom regime is permitted. The role play by the $\Gamma$ term is to turn the minimum value of $\omega(a)$ a exponential function of time. If dark energy has a source of particles, $\Gamma>0$, $\omega(a)_{min}$ is a exponential growth, diverging from $-1$ as times goes by. However, if the fluid suffers from particle destruction processes, then $\omega(a)_{min}$ is a exponential decay and therefore converges to $-1$. This behavior is represented in the theoretical graphs of Fig.(\ref{fig1})  and Fig.(\ref{fig2}). 

\begin{figure}[t]
    \centering
    \includegraphics[width=0.71\textwidth]{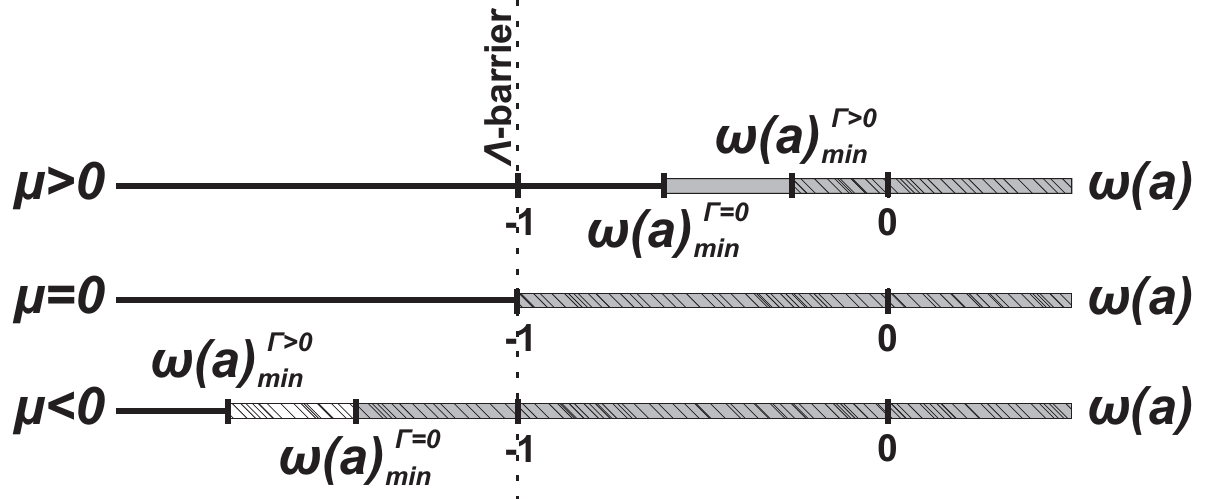}
    \caption{\justifying Comparative scheme for the $\omega(a)$ interval when $\Gamma=0$ (continuous part) and $\Gamma>0$ (hatched part) for a given instant of time. The presence of a particle source causes $\omega(a)$ to move away from the $\Lambda$ barrier independent of the sign of the chemical potential. Image based on \cite{silva2013}.}
    \label{fig3}
\end{figure}

\begin{figure}
    \centering
    \includegraphics[width=0.71\textwidth]{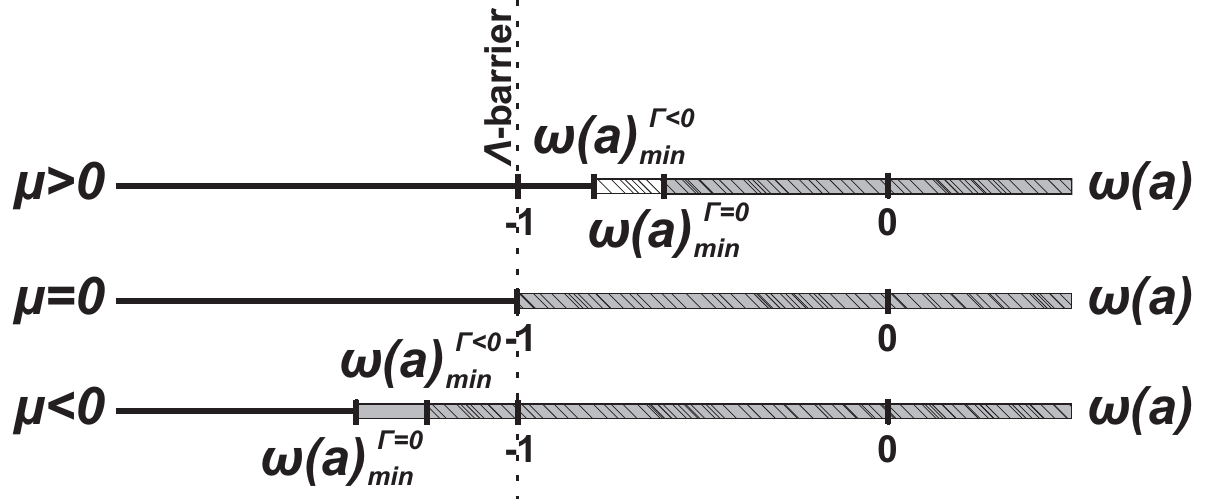}
    \caption{\justifying Comparative scheme for the $\omega(a)$ interval when $\Gamma=0$ (continuous part) and $\Gamma<0$ (hatched part) for a given instant of time. The presence of a particle sink causes $\omega(a)$ to move closer to the $\Lambda$ barrier independent of the sign of the chemical potential. Image based on \cite{silva2013}.}
    \label{fig4}
\end{figure}

Note that if we have a particle source in the fluid, the values of $\omega(a)_{min}$ will gradually increase, becoming more positive (positive chemical potential) or more negative (negative chemical potential). We can also have no changing at all (null chemical potential). For a given instant of time, Fig.(\ref{fig3}) compares the old interval of values for the EoS parameter to the new one after the inclusion of $\Gamma>0$. We observe that particle creation makes dark energy distance itself to the cosmological constant. On the other hand, if we have a sink in the fluid, the minimum value of the EoS parameter will gradually decrease, but with an asymptote at $\omega(a)_{min}=-1$. So, when dark energy suffers with particle destruction processes, it tends to the cosmological constant. In Fig.(\ref{fig4}) we have a scheme analogous to Fig.(\ref{fig3}) for $\Gamma<0$.

Since the cosmological constant still is the best explanation for cosmic acceleration \cite{carroll, peebles}, our model leads us to conclude that it is more likely to find a sink of particles than a source in the dark energy fluid. As we can see from the above development, this does not break either the second law of thermodynamics or any other.
\section{Conclusions} \label{conclusions}

In this work, we proposed a new thermodynamic model for dark energy that generalize the one discussed in \cite{silva2013} by considering processes of creation or destruction of particles. Despite the dark energy density evolution and the temperature evolution law remains intact, we derive new expressions for the evolution of particle density and entropy density, showing that they reduce to the standards equations if $\Gamma=0$. We also obtained new thermodynamics constraints which shows that $\omega(a)_{min}$ starts to be a exponential function of time in the presence of a particle source or sink, with this behavior being evidenced in Figs.(\ref{fig1}, \ref{fig2}). In particular, we noticed that if dark energy suffers from particle destruction processes, then the fluid will eventually behaves like the cosmological constant regardless the sign of its chemical potential. Contrary to what one might be expected based on \cite{gunzig, prigogine1988, prigogine1989}, our arguments leading to this conclusion do not go against what the laws of thermodynamics say. This is a unprecedented result that our model predicts. Not only that, but all of them goes back to \cite{silva2013} if there is no particle source or sink. 

Some questions, however, may emerge from this analysis. The more immediate one is about the nature of the particles created or destroyed in the dark energy fluid. Since the nature of this exotic component itself is not known so far, it is difficult to give a straight answer. One may look for geometric justifications \cite{biswas2002} or a possible interaction with dark matter - some universe models considers a cosmos suffering with creation of dark matter particles \cite{ema2018, herring2020, cardenas2020.1,cardenas2020.2}. Not only that, but it is of fundamental importance to give a definite form to the $\Gamma$ term: if it is a constant, we will have a usual exponential growth for $\Gamma>0$ or decay for $\Gamma<0$, which is a very simple scenario; if it depends on $a$ in some order, it is not that trivial. Finally, it is necessary to confront this study with the newest observational data to verify if our theoretical predictions are consistent. In this sense, we expect to evolve this initial model to a more fulfilled one and (try) to give a good enough answer to those questions, formulating new ones in the process.
\section*{Acknowledgments}

We would like to thanks the Fundação de Amparo à Ciência e Tecnologia do Estado de Pernambuco (FACEPE) for all the financial support provided. 
\bibliographystyle{model1-num-names} 
\bibliography{paper}

\begin{thebibliography}{38}
\expandafter\ifx\csname natexlab\endcsname\relax\def\natexlab#1{#1}\fi
\providecommand{\url}[1]{\texttt{#1}}
\providecommand{\href}[2]{#2}
\providecommand{\path}[1]{#1}
\providecommand{\DOIprefix}{doi:}
\providecommand{\ArXivprefix}{arXiv:}
\providecommand{\URLprefix}{URL: }
\providecommand{\Pubmedprefix}{pmid:}
\providecommand{\doi}[1]{\href{http://dx.doi.org/#1}{\path{#1}}}
\providecommand{\Pubmed}[1]{\href{pmid:#1}{\path{#1}}}
\providecommand{\bibinfo}[2]{#2}
\ifx\xfnm\relax \def\xfnm[#1]{\unskip,\space#1}\fi
\bibitem[{Riess et~al.(1998)}]{riess}
\bibinfo{author}{A.~G. Riess}, et~al.,
\newblock \bibinfo{title}{Observational evidence from supernovae for an accelerating universe and a cosmological constant},
\newblock \bibinfo{journal}{Astron. J.} \bibinfo{volume}{116} (\bibinfo{year}{1998}) \bibinfo{pages}{1009--1038}.
\bibitem[{Perlmutter et~al.(1999)}]{perlmutter}
\bibinfo{author}{S.~Perlmutter}, et~al.,
\newblock \bibinfo{title}{{Measurements of $\Omega$ and $\Lambda$ from 42 high-redshift supernovae}},
\newblock \bibinfo{journal}{Astrophys. J.} \bibinfo{volume}{517} (\bibinfo{year}{1999}) \bibinfo{pages}{565--586}.
\bibitem[{Aghanim et~al.(2020)}]{aghanim}
\bibinfo{author}{N.~Aghanim}, et~al.,
\newblock \bibinfo{title}{{Planck 2018 results-VI. Cosmological parameters}},
\newblock \bibinfo{journal}{Astron. Astrophys.} \bibinfo{volume}{641} (\bibinfo{year}{2020}) \bibinfo{pages}{A6}.
\bibitem[{Weinberg(1972)}]{weinberg1972}
\bibinfo{author}{S.~Weinberg}, \bibinfo{title}{Gravitation and cosmology: principles and applications of the general theory of relativity}, \bibinfo{edition}{first} ed., \bibinfo{publisher}{John Wiley and Sons}, \bibinfo{address}{New York}, \bibinfo{year}{1972}.
\bibitem[{Weinberg(2008)}]{weinberg2008}
\bibinfo{author}{S.~Weinberg}, \bibinfo{title}{Cosmology}, \bibinfo{edition}{first} ed., \bibinfo{publisher}{OUP Oxford}, \bibinfo{address}{New York}, \bibinfo{year}{2008}.
\bibitem[{Carroll(2001)}]{carroll}
\bibinfo{author}{S.~M. Carroll},
\newblock \bibinfo{title}{The cosmological constant},
\newblock \bibinfo{journal}{Living Rev. Relativ.} \bibinfo{volume}{4} (\bibinfo{year}{2001}) \bibinfo{pages}{1--56}.
\bibitem[{Peebles and Ratra(2003)}]{peebles}
\bibinfo{author}{P.~J.~E. Peebles}, \bibinfo{author}{B.~Ratra},
\newblock \bibinfo{title}{The cosmological constant and dark energy},
\newblock \bibinfo{journal}{Rev. Mod. Phys.} \bibinfo{volume}{75} (\bibinfo{year}{2003}) \bibinfo{pages}{559--606}.
\bibitem[{Lima and Alcaniz(2004)}]{lima2004}
\bibinfo{author}{J.~A.~S. Lima}, \bibinfo{author}{J.~S. Alcaniz},
\newblock \bibinfo{title}{Thermodynamics, spectral distribution and the nature of dark energy},
\newblock \bibinfo{journal}{Phys. Lett. B} \bibinfo{volume}{600} (\bibinfo{year}{2004}) \bibinfo{pages}{191--196}.
\bibitem[{González-Díaz(2003)}]{gonzalez2003}
\bibinfo{author}{P.~F. González-Díaz},
\newblock \bibinfo{title}{You need not be afraid of phantom energy},
\newblock \bibinfo{journal}{Phys. Rev. D} \bibinfo{volume}{68} (\bibinfo{year}{2003}) \bibinfo{pages}{021303}.
\bibitem[{Sami and Toporensky(2004)}]{sami}
\bibinfo{author}{M.~Sami}, \bibinfo{author}{A.~Toporensky},
\newblock \bibinfo{title}{Phantom field and the fate of the universe},
\newblock \bibinfo{journal}{Mod. Phys. Lett. A} \bibinfo{volume}{19} (\bibinfo{year}{2004}) \bibinfo{pages}{1509--1517}.
\bibitem[{González-Díaz and Sigüenza(2004{\natexlab{a}})}]{gonzalez2004.1}
\bibinfo{author}{P.~F. González-Díaz}, \bibinfo{author}{C.~L. Sigüenza},
\newblock \bibinfo{title}{The fate of black holes in an accelerating universe},
\newblock \bibinfo{journal}{Phys. Lett. B} \bibinfo{volume}{589} (\bibinfo{year}{2004}{\natexlab{a}}) \bibinfo{pages}{78--82}.
\bibitem[{González-Díaz and Sigüenza(2004{\natexlab{b}})}]{gonzalez2004.2}
\bibinfo{author}{P.~F. González-Díaz}, \bibinfo{author}{C.~L. Sigüenza},
\newblock \bibinfo{title}{Phantom thermodynamics},
\newblock \bibinfo{journal}{Nucl. Phys. B} \bibinfo{volume}{697} (\bibinfo{year}{2004}{\natexlab{b}}) \bibinfo{pages}{363--386}.
\bibitem[{Babichev et~al.(2004)Babichev, Dokuchaev, and Eroshenko}]{babichev}
\bibinfo{author}{E.~Babichev}, \bibinfo{author}{V.~Dokuchaev}, \bibinfo{author}{Y.~Eroshenko},
\newblock \bibinfo{title}{Black hole mass decreasing due to phantom energy accretion},
\newblock \bibinfo{journal}{Phys. Rev. Lett.} \bibinfo{volume}{93} (\bibinfo{year}{2004}) \bibinfo{pages}{021102}.
\bibitem[{Myung(2009)}]{myung}
\bibinfo{author}{Y.~S. Myung},
\newblock \bibinfo{title}{On phantom thermodynamics with negative temperature},
\newblock \bibinfo{journal}{Phys. Lett. B} \bibinfo{volume}{671} (\bibinfo{year}{2009}) \bibinfo{pages}{216--218}.
\bibitem[{Saridakis et~al.(2009)Saridakis, González-Díaz, and Sigüenza}]{saridakis}
\bibinfo{author}{E.~N. Saridakis}, \bibinfo{author}{P.~F. González-Díaz}, \bibinfo{author}{C.~L. Sigüenza},
\newblock \bibinfo{title}{Unified dark energy thermodynamics: varying $\omega$ and the $-1$-crossing},
\newblock \bibinfo{journal}{Class. Quantum Gravity} \bibinfo{volume}{26} (\bibinfo{year}{2009}) \bibinfo{pages}{165003}.
\bibitem[{Nojiri and Odintsov(2003{\natexlab{a}})}]{nojiri2003.1}
\bibinfo{author}{S.~Nojiri}, \bibinfo{author}{S.~D. Odintsov},
\newblock \bibinfo{title}{Quantum de sitter cosmology and phantom matter},
\newblock \bibinfo{journal}{Phys. Lett. B} \bibinfo{volume}{562} (\bibinfo{year}{2003}{\natexlab{a}}) \bibinfo{pages}{147--152}.
\bibitem[{Nojiri and Odintsov(2003{\natexlab{b}})}]{nojiri2003.2}
\bibinfo{author}{S.~Nojiri}, \bibinfo{author}{S.~D. Odintsov},
\newblock \bibinfo{title}{Effective equation of state and energy conditions in phantom/tachyon inflationary cosmology perturbed by quantum effects},
\newblock \bibinfo{journal}{Phys. Lett. B} \bibinfo{volume}{571} (\bibinfo{year}{2003}{\natexlab{b}}) \bibinfo{pages}{1--10}.
\bibitem[{Cline et~al.(2004)Cline, Jeon, and Moore}]{cline}
\bibinfo{author}{J.~M. Cline}, \bibinfo{author}{S.~Jeon}, \bibinfo{author}{G.~D. Moore},
\newblock \bibinfo{title}{The phantom menaced: Constraints on low-energy effective ghosts},
\newblock \bibinfo{journal}{Phys. Rev. D} \bibinfo{volume}{70} (\bibinfo{year}{2004}) \bibinfo{pages}{043543}.
\bibitem[{Samart and Gumjudpai(2007)}]{samart}
\bibinfo{author}{D.~Samart}, \bibinfo{author}{B.~Gumjudpai},
\newblock \bibinfo{title}{Phantom field dynamics in loop quantum cosmology},
\newblock \bibinfo{journal}{Phys. Rev. D} \bibinfo{volume}{76} (\bibinfo{year}{2007}) \bibinfo{pages}{043514}.
\bibitem[{Lima and Pereira(2008)}]{lima2008}
\bibinfo{author}{J.~A.~S. Lima}, \bibinfo{author}{S.~H. Pereira},
\newblock \bibinfo{title}{Chemical potential and the nature of dark energy: The case of a phantom field},
\newblock \bibinfo{journal}{Phys. Rev. D} \bibinfo{volume}{78} (\bibinfo{year}{2008}) \bibinfo{pages}{083504}.
\bibitem[{Pereira and Lima(2008)}]{pereira}
\bibinfo{author}{S.~H. Pereira}, \bibinfo{author}{J.~A.~S. Lima},
\newblock \bibinfo{title}{On phantom thermodynamics},
\newblock \bibinfo{journal}{Phys. Lett. B} \bibinfo{volume}{669} (\bibinfo{year}{2008}) \bibinfo{pages}{266--270}.
\bibitem[{Barboza~Jr and Alcaniz(2008)}]{barboza}
\bibinfo{author}{E.~M. Barboza~Jr}, \bibinfo{author}{J.~S. Alcaniz},
\newblock \bibinfo{title}{A parametric model for dark energy},
\newblock \bibinfo{journal}{Phys. Lett. B} \bibinfo{volume}{666} (\bibinfo{year}{2008}) \bibinfo{pages}{415--419}.
\bibitem[{Silva et~al.(2012)Silva, Goncalves, Alcaniz, and Silva}]{silva2012}
\bibinfo{author}{R.~Silva}, \bibinfo{author}{R.~S. Goncalves}, \bibinfo{author}{J.~S. Alcaniz}, \bibinfo{author}{H.~H.~B. Silva},
\newblock \bibinfo{title}{Thermodynamics and dark energy},
\newblock \bibinfo{journal}{Astron. Astrophys.} \bibinfo{volume}{537} (\bibinfo{year}{2012}) \bibinfo{pages}{A11}.
\bibitem[{Silva et~al.(2013)Silva, Silva, Gonçalves, Zhu, and Alcaniz}]{silva2013}
\bibinfo{author}{H.~H.~B. Silva}, \bibinfo{author}{R.~Silva}, \bibinfo{author}{R.~S. Gonçalves}, \bibinfo{author}{Z.-H. Zhu}, \bibinfo{author}{J.~S. Alcaniz},
\newblock \bibinfo{title}{General treatment for dark energy thermodynamics},
\newblock \bibinfo{journal}{Phys. Rev. D} \bibinfo{volume}{88} (\bibinfo{year}{2013}) \bibinfo{pages}{127302}.
\bibitem[{Gunzig et~al.(1987)Gunzig, Geheniau, and Prigogine}]{gunzig}
\bibinfo{author}{E.~Gunzig}, \bibinfo{author}{J.~Geheniau}, \bibinfo{author}{I.~Prigogine},
\newblock \bibinfo{title}{Entropy and cosmology},
\newblock \bibinfo{journal}{Nature} \bibinfo{volume}{330} (\bibinfo{year}{1987}) \bibinfo{pages}{621--624}.
\bibitem[{Prigogine et~al.(1988)Prigogine, Géhéniau, Gunzig, and Nardone}]{prigogine1988}
\bibinfo{author}{I.~Prigogine}, \bibinfo{author}{J.~Géhéniau}, \bibinfo{author}{E.~Gunzig}, \bibinfo{author}{P.~Nardone},
\newblock \bibinfo{title}{Thermodynamics of cosmological matter creation},
\newblock \bibinfo{journal}{Proc. Natl. Acad. Sci. USA} \bibinfo{volume}{85} (\bibinfo{year}{1988}) \bibinfo{pages}{7428--7432}.
\bibitem[{Prigogine(1989)}]{prigogine1989}
\bibinfo{author}{I.~Prigogine},
\newblock \bibinfo{title}{Thermodynamics and cosmology},
\newblock \bibinfo{journal}{Int. J. Theor. Phys.} \bibinfo{volume}{28} (\bibinfo{year}{1989}) \bibinfo{pages}{927--933}.
\bibitem[{Calvão et~al.(1992)Calvão, Lima, and Waga}]{calvao}
\bibinfo{author}{M.~O. Calvão}, \bibinfo{author}{J.~A.~S. Lima}, \bibinfo{author}{I.~Waga},
\newblock \bibinfo{title}{On the thermodynamics of matter creation in cosmology},
\newblock \bibinfo{journal}{Phys. Lett. A} \bibinfo{volume}{162} (\bibinfo{year}{1992}) \bibinfo{pages}{223--226}.
\bibitem[{Lima and Germano(1992)}]{lima1992}
\bibinfo{author}{J.~A.~S. Lima}, \bibinfo{author}{A.~S.~M. Germano},
\newblock \bibinfo{title}{On the equivalence of bulk viscosity and matter creation},
\newblock \bibinfo{journal}{Phys. Lett. A} \bibinfo{volume}{170} (\bibinfo{year}{1992}) \bibinfo{pages}{373--378}.
\bibitem[{Zimdahl and Pavón(1993)}]{zimdahl1993}
\bibinfo{author}{W.~Zimdahl}, \bibinfo{author}{D.~Pavón},
\newblock \bibinfo{title}{Cosmology with adiabatic matter creation},
\newblock \bibinfo{journal}{Phys. Lett. A} \bibinfo{volume}{176} (\bibinfo{year}{1993}) \bibinfo{pages}{57--61}.
\bibitem[{Lima et~al.(1996)Lima, Germano, and Abramo}]{lima1996}
\bibinfo{author}{J.~A.~S. Lima}, \bibinfo{author}{A.~S.~M. Germano}, \bibinfo{author}{L.~R.~W. Abramo},
\newblock \bibinfo{title}{{FRW-type cosmologies with adiabatic matter creation}},
\newblock \bibinfo{journal}{Phys. Rev. D} \bibinfo{volume}{53} (\bibinfo{year}{1996}) \bibinfo{pages}{4287--4297}.
\bibitem[{Zimdahl(1996)}]{zimdahl1996}
\bibinfo{author}{W.~Zimdahl},
\newblock \bibinfo{title}{Bulk viscous cosmology},
\newblock \bibinfo{journal}{Phys. Rev. D} \bibinfo{volume}{53} (\bibinfo{year}{1996}) \bibinfo{pages}{5483--5493}.
\bibitem[{Reichl(2016)}]{reichl}
\bibinfo{author}{L.~Reichl}, \bibinfo{title}{A modern course in statistical physics}, \bibinfo{edition}{fourth} ed., \bibinfo{publisher}{John Wiley and Sons}, \bibinfo{address}{New York}, \bibinfo{year}{2016}.
\bibitem[{Biswas et~al.(2002)Biswas, Shaw, and Misra}]{biswas2002}
\bibinfo{author}{S.~Biswas}, \bibinfo{author}{A.~Shaw}, \bibinfo{author}{P.~Misra},
\newblock \bibinfo{title}{Particle production in expanding spacetime},
\newblock \bibinfo{journal}{Gen. Relativ. Gravit.} \bibinfo{volume}{34} (\bibinfo{year}{2002}) \bibinfo{pages}{665--678}.
\bibitem[{Ema et~al.(2018)Ema, Nakayama, and Tang}]{ema2018}
\bibinfo{author}{Y.~Ema}, \bibinfo{author}{K.~Nakayama}, \bibinfo{author}{Y.~Tang},
\newblock \bibinfo{title}{Production of purely gravitational dark matter},
\newblock \bibinfo{journal}{J. High Energy Phys.} \bibinfo{volume}{2018} (\bibinfo{year}{2018}) \bibinfo{pages}{1--20}.
\bibitem[{Herring and Boyanovsky(2020)}]{herring2020}
\bibinfo{author}{N.~Herring}, \bibinfo{author}{D.~Boyanovsky},
\newblock \bibinfo{title}{Gravitational production of nearly thermal fermionic dark matter},
\newblock \bibinfo{journal}{Phys. Rev. D} \bibinfo{volume}{101} (\bibinfo{year}{2020}) \bibinfo{pages}{123522}.
\bibitem[{Cárdenas et~al.(2020{\natexlab{a}})Cárdenas, Cruz, Lepe, Nojiri, and Odintsov}]{cardenas2020.1}
\bibinfo{author}{V.~H. Cárdenas}, \bibinfo{author}{M.~Cruz}, \bibinfo{author}{S.~Lepe}, \bibinfo{author}{S.~Nojiri}, \bibinfo{author}{S.~D. Odintsov},
\newblock \bibinfo{title}{Challenging matter creation models in the phantom divide},
\newblock \bibinfo{journal}{Phys. Rev. D} \bibinfo{volume}{101} (\bibinfo{year}{2020}{\natexlab{a}}) \bibinfo{pages}{083530}.
\bibitem[{Cárdenas et~al.(2020{\natexlab{b}})Cárdenas, Cruz, and Lepe}]{cardenas2020.2}
\bibinfo{author}{V.~H. Cárdenas}, \bibinfo{author}{M.~Cruz}, \bibinfo{author}{S.~Lepe},
\newblock \bibinfo{title}{Cosmic expansion with matter creation and bulk viscosity},
\newblock \bibinfo{journal}{Phys. Rev. D} \bibinfo{volume}{102} (\bibinfo{year}{2020}{\natexlab{b}}) \bibinfo{pages}{123543}.

\end{thebibliography}
\end{document}